# POLYMER MELT VISCOSITY

A. N. Yakunin[1]

Karpov Institute of Physical Chemistry, ul.Vorontsovo Pole 10, Moscow 103064, Russia

---

[1] E-mail: yakunin@cc.nifhi.ac.ru

**Abstract:** The research is important for a molecular theory of liquid and has a wide interest as an example solving the problem when dynamic parameters of systems can be indirectly connected with their equilibrium properties. In frameworks of the reptation model the power law with the 3.5-exponent for the melt viscosity relation to the molecular weight of flexible-chain polymer is obtained. In order to find the exponent close to experimental values it should be taken into account the rotation vibration precession motion of chain ends with respect to the polymer melt flow direction.

# 1. Introduction

In present, one may describe the polymer melt viscosity dependence on the polymerisation degree (the number of chain monomers per one macromolecule) in terms of the reptation model [1-3] suggested by P.-G. de Gennes in 1971. The reptation theory explains why the power law for the melt viscosity relation to the molecular weight of polymer can be observed. However, experimental data show the strong discrepancy with the theory (the exponent is equal to 3.3-3.4 or even is out of the range as distinct from the theoretical value which is equal to 3). Usually, in order to obtain better agreement with experiment one may assume that the length of the tube, created by entanglement chains and along that a macromolecule can crawl making the reptation motion, can be subjected to fluctuations [2]. Wool [4] has also derived an analogous result using a different concept. Other attempts have been aimed modifying the dynamic equations [5, 6]. They continue up to now. More complex systems such as phase-separate polymer solutions, blends, block and graft copolymer mesophases and other fluids are suppose to be applicable to studying by similar techniques [7]. However, in such approaches the number of monomers between neighbour entanglements along the chain, $N_e$, is introduced phenomenologically or ignored completely. At last, the recent Monte Carlo results [8] have shown that reptation motions prevail in melts where entropic trapping is absent in contrast to swollen gels [9]. Thus, the questions dealing with polymer melt viscosity and entanglements have been extensively discussed during the past decade [3, 4, 7, 8, 10]. The fact claims that the problems are open in the present time although the entanglement concentration has been already estimated (see, for example, [4]).

The aim of the article is to construct the solution by such way which could enable to achieve a good accordance with the experimental data in frameworks of the reptation model without the use of no additional simplifying assumptions.

The structure of the article is following. We will recall the main principles of the reptation theory [1-3] and try to understand which of ways can be fitted to modify the theory. We will see that we should introduce the notation of a mechanical field. In that case, a chain seems can "swell" and the vector connecting the chain ends can rotate (Figure 1) making a vibration precession motion with respect to the direction of polymer melt flow. Finally, we will obtain a good agreement with the experimental data by help of the suggested changed reptation theory. A physical transparency of concepts forming its background makes very attractive the application of the approach in understanding and explaining the nature of the energy dissipation in liquids.

## 2. Results and discussion

Let us consider the reptation theory [1]. A macromolecule of flexible-chain polymer melt crawls along the tube under the influence of an applied constant force, f, with the resulting velocity, v. Let $S_0$ be the cross-section area of the tube. Assuming $S_0^{1/2} = aN_e^{1/2}$ where a is the diameter of the monomer (diameter of the chain or the lattice constant) we can find the expression for the length of the tube, L (in units of the diameter of the tube):

$$L = S_0^{1/2} N/N_e = aN/N_e^{1/2}. \qquad (1)$$

Here, N is the polymerisation degree. It should be underlined that we are not going to write numerical constants in our formulae because we only want to obtain the true exponent for the melt viscosity dependence on the molecular weight of polymer.

The mobility of the chain along the tube, $\mu_t = 1/\zeta_{fr}$, may be defined as

$$\mu_t = v/f \qquad (2)$$

where $\zeta_{fr}$ is the friction coefficient. Finally, using (1) and (2) we can find the time of relaxation (so-called the maximum time of relaxation), $\tau$:

$$\tau = L^2/D_t = (aN/N_e^{1/2})^2/(vTN^{-1}/f_1) = \tau_1 N^3/N_e \qquad (3)$$

where $D_t = \mu_t T = vTN^{-1}/f_1$ is the diffusion coefficient of the chain along the tube, $f_1$ is the friction force per one monomer, T is the temperature expressed in energy units, $\tau_1 = a^2/D_1$ and $D_1$ are the time of relaxation and the diffusion coefficient typical for low-molecular-weight liquids, respectively. Thus, from (3) one may draw the following conclusion: the polymer melt viscosity

$$\eta \sim E\tau \quad (4)$$

is proportional to the polymerisation degree to the third power. Here,

$$E \sim Ta^{-3}/N_e \quad (5)$$

is the elastic modulus of the fluctuation network of polymer melt.

Now, we are going to suggest the main scaling arguments for our model. We do not want to change the formula (1), consequently, we should find the mechanism decreasing the mobility of the reptation motion. We are going to look for an additional force responsible for the appropriate increase of viscosity.

Let us consider the such mechanical field for that the frequency, $\omega$, of the rotation vibration motion of the chain ends with respect to the polymer melt flow direction is much less than the reciprocal value of $\tau$

$$\varphi = \omega\tau \ll 1. \quad (6)$$

Then the arising phase difference, $\varphi$, is connected with the moment, M, which tends to return the chain ends in their initial position, by help of the expression

$$M \sim T\varphi/g_D. \quad (7)$$

Here, $g_D \sim r^{-1}$ is the Debye pair correlation function [1-3], r is the end-to-end distance and the scale at which the field significantly changes, $r \sim aN^{1/2}$ for ideal chains [1-3]. The ratio $M/\omega$ from (6, 7) can be expressed through the true viscosity, $\eta_{tr}$:

$$\eta_{tr} = \text{const}(M/\omega) = \text{const}(T\tau r). \quad (8)$$

If we will regard the constant equal to $a^{-4}N_e^{-1}$ then the expression (8) will look like (4) with E from (5) and the true relaxation time, $\tau_{tr}$

$$\tau_{tr} \sim \tau \, N^{1/2}. \tag{9}$$

Thus, the maximum time of relaxation (9) increases resulting in the hydrodynamic effect of the velocity of a ball-shaped body, rotating effectively in a high-molecular-weight liquid with the viscosity $\eta$, on the friction force f. A characteristic size of the body is r. It means as we think that there exists an fluctuation attraction between the chain ends since the end monomers can locate into an interaction potential differing from that of other monomers. The mechanism of the interaction in polymer solutions was described elsewhere [3].

We must note that although all motions associated with the chain ends are fast at small scales of the order of several atom radii, the reduction in overall chain mobility is due to a mechanism which 'inhibits' the motions of the ends at large scales (of the order of the end-to-end distance).

The rotation vibration motion can not be observed at macroscopic scales due to decreasing the field with increasing the distance. More accurate accordance with experimental data can be not achieved owing to the interaction potential at short distances depends on chemical nature of polymers.

We have also found that an non-equilibrium parameter of polymer melts such as the viscosity is indirectly connected with the equilibrium pair correlation function due to the finite length of macromolecules. We can indeed neglect the fluctuation attraction of the chain ends, supposing it small, for long enough chains.

**Acknowledgements**

Thanks are expressed to Professor A.R.Khokhlov and to Professor I.Ya.Erukhimovich from Moscow State University for useful discussions, and also to Dr. A.V.Mironov for technical help in preparing the manuscript.

The author is thankful to Russian Foundation for Basic Research (Grant No 01-03-32225) for financial support.

**Figure caption**

Figure 1. A macromolecule crawling into entanglement network (open circles) can make the rotation vibration motion (shown by vectors) resulting in increasing viscosity. Dotted curve designates the initial position of the tube, and solid curve - its final position.

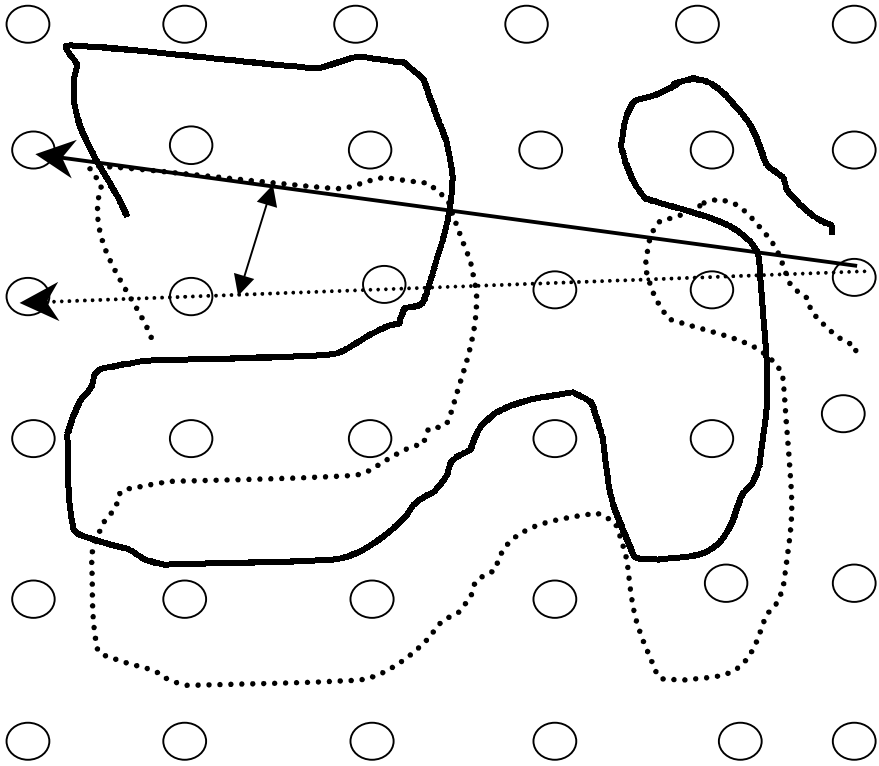